\documentclass[11pt]{article}
\usepackage[utf8]{inputenc}
\usepackage[T1]{fontenc}
\usepackage{lmodern}
\usepackage[margin=1in]{geometry}
\usepackage{amsmath,amssymb,bm}
\usepackage{booktabs}
\usepackage{tabularx}
\usepackage{graphicx}
\usepackage{caption}
\usepackage{subcaption}
\usepackage{enumitem}
\usepackage{placeins}
\usepackage{microtype}
\usepackage{tikz}
\usepackage{xspace}
\usetikzlibrary{positioning}
\usepackage[numbers,sort&compress]{natbib}
\usepackage[hidelinks]{hyperref}

\title{Accuracy Analysis of VLBI Universal Time Measurement Based on a GNSS Single-Station Regional Ionospheric Model}
\author{Xu-chong Duan$^{1,2,3,4}$, Dang Yao$^{1,2}$, Yuan-wei Wu$^{1,2,3,4}$, Zhe Zhang$^{1,2}$, Jia Liu$^{1,2}$, and Xu-hai Yang$^{1,2,3,4}$\\[3pt]
\small $^1$National Time Service Center, Chinese Academy of Sciences, Xi'an 710600, China\\
\small $^2$Key Laboratory of Time Reference and Applications, Chinese Academy of Sciences, Xi'an 710600, China\\
\small $^3$University of Chinese Academy of Sciences, Beijing 100049, China\\
\small $^4$School of Astronomy and Space Science, University of Chinese Academy of Sciences, Beijing 100049, China}
\date{}

\newcommand{\UT}{UT1\xspace}

\newcommand{\VTEC}{\mathrm{VTEC}}
\newcommand{\STEC}{\mathrm{STEC}}

\begin{document}
\maketitle

\begin{abstract}
Universal Time (\UT) is a key parameter characterizing Earth's rotation, and very long baseline interferometry (VLBI) is the mainstream technique for measuring \UT. To address the limitations in the timeliness and accuracy of existing global ionospheric models for single-frequency VLBI \UT measurements, we construct a single-station regional ionospheric model using GNSS data from the VLBI stations on the Jilin--Kashi baseline. We apply this model to VLBI observations and compare its correction performance with that of a global predictive model and a global post-processed model. The results show that the line-of-sight ionospheric delays and baseline corrections calculated with the single-station regional model have precision close to that of the global post-processed model and are substantially better than those of the global predictive model. After correction with the single-station regional model, the derived \UT values differ from the US Naval Observatory (USNO) reference values by a mean bias of $-15.6\,\mu$s and an RMS deviation of $82.3\,\mu$s, both better than the results obtained with the other two model classes. A single-station regional ionospheric model constructed independently from GNSS data available at VLBI stations can effectively correct single-frequency VLBI observations and support quasi-real-time high-precision \UT measurements. It therefore has important value for improving the timeliness of independent \UT products.
\end{abstract}


\noindent\textbf{Article DOI:} 10.13875/j.issn.1674-0637.2024-01-0060-11

\noindent\textbf{Suggested citation:} Duan, X.-c., Yao, D., Wu, Y.-w., et al. Accuracy Analysis of VLBI Universal Time Measurement Based on a GNSS Single-Station Regional Ionospheric Model. \emph{Journal of Time and Frequency}, 2026, 47(1): 1--19.

\section{Introduction}
Universal Time (UT) is a time scale defined by the rotation of the Earth \citep{ref1}. In 2000, the International Astronomical Union (IAU) adopted the IAU2000 Earth rotation model, which redefined the UT1 parameter so that it is linearly related to the Earth Rotation Angle (ERA) between the terrestrial and celestial reference frames. Together with the polar-motion parameters (PMX/PMY) and celestial pole offsets (dX/dY), \UT forms the Earth orientation parameters (EOP) that describe the state of Earth's rotation. These parameters are indispensable basic data for satellite navigation, deep-space exploration, geodesy, and geodynamics.

At present, VLBI is the only space-geodetic technique capable of stably monitoring \UT over the long term \citep{ref2}. The technique uses several radio telescopes distributed in different regions to observe distant extragalactic compact radio sources synchronously, determines the signal arrival-time differences at the stations, and then estimates \UT through least-squares parameter estimation. Thus, the measurement and calibration accuracy of the baseline delay directly determine the accuracy of the final \UT estimate.

The path from a radio source to a receiving antenna is not an ideal vacuum. During propagation through interstellar space, the signal is affected by charged interstellar media; after entering the Earth's atmosphere, it experiences refractive delays in the ionosphere and troposphere. In addition, Earth rotation continuously changes the geometric path between the signal and the stations. Antenna structural deformation, local-oscillator clock offsets and frequency drift, and electronic delays in receiving equipment and transmission cables also perturb the signal propagation time. The observed delay at the receiver therefore mainly consists of geometric, tropospheric, ionospheric, and antenna-deformation delays, among others \citep{ref3}. The observation equation can be written as

\begin{equation}
\tau^{\mathrm{obs}}_{ij}=\tau^{\mathrm{geom}}_{ij}+\tau^{\mathrm{iono}}_{ij}+\tau^{\mathrm{trop}}_{ij}+\tau^{\mathrm{clk}}_{ij}+\tau^{\mathrm{inst}}_{ij}+\tau^{\mathrm{ant}}_{ij}+\tau^{\mathrm{grav}}_{ij}+\epsilon,
\label{eq:vlbi}
\end{equation}
where the terms denote, respectively, the observed group-delay difference, geometric-delay difference, ionospheric-delay difference, tropospheric-delay difference, station-clock difference, instrumental-delay difference, antenna-structural-deformation delay difference, and other gravitational or related delay differences between the two stations; $\epsilon$ is random noise. In VLBI \UT determination, the vacuum geometric delay is the central quantity to be estimated. It is determined by the geometry of the stations, radio source, and Earth and carries the geodynamic information. Ionospheric delay is caused by free electrons along the signal path and must be removed before parameter estimation. Common corrections either use dual-frequency (for example, S/X-band) observations to eliminate the ionospheric delay or introduce an external ionospheric model. Tropospheric delay is caused by neutral gases and water vapour, depends on station position, observing direction, and observing time, and is usually estimated as a parameter. Station-clock delay represents the offset and frequency drift between each station's independent oscillator and the reference time scale and is also generally estimated. Instrumental delay mainly refers to the electronic delay in the receiving and transmission chain and is accurately removed through observations of calibration sources. Antenna-deformation delay is mainly caused by gravity, wind loading, and temperature effects and can be corrected with a telescope model. Other delays include effects from relativity, solid Earth tides, ocean tides, polar tides, and ocean loading and are generally corrected with precise models such as IAU2000.

\begin{figure}[tbp]
\centering
\begin{tikzpicture}[scale=0.86, every node/.style={font=\small}]
  \draw[thick] (-4.2,1.35) .. controls (-2.2,2.35) and (2.2,2.35) .. (4.2,1.35);
  \draw[thick] (-3.8,2.25) .. controls (-2.0,3.00) and (2.0,3.00) .. (3.8,2.25);
  \fill[gray!35] (-3.25,0.10) .. controls (-1.7,0.95) and (1.7,0.95) .. (3.25,0.10)
    .. controls (1.7,-0.55) and (-1.7,-0.55) .. cycle;
  \coordinate (O) at (0,-3.0);
  \coordinate (S) at (0,0.25);
  \coordinate (P) at (1.85,1.10);
  \coordinate (Q) at (3.75,2.38);
  \draw[thick] (O) -- (S);
  \draw[thick] (O) -- (P);
  \draw[thick] (S) -- (Q);
  \draw[thick,->] (S) -- (0,2.28);
  \draw (P) -- ++(0.55,0.85);
  \draw (P) ++(0.55,0.85) -- (Q);
  \fill (S) circle (2pt);
  \fill (P) circle (2pt);
  \fill (Q) circle (2pt);
  \node[above] at (0,2.32) {Ionosphere};
  \node[right] at (Q) {Radio source};
  \node[right] at (P) {Pierce point};
  \node[left] at (S) {Station};
  \node[right] at (2.95,-2.95) {Station};
  \node[left] at (0,-1.45) {$R$};
  \node[left] at (0,1.35) {$H$};
  \node[right] at (0.25,-2.45) {$\alpha$};
  \node[right] at (0.16,0.55) {$z$};
  \node[right] at (2.05,1.72) {$z'$};
\end{tikzpicture}
\caption{Geometric relationship between VTEC and STEC during signal propagation.}
\label{fig:vtec-stec}
\end{figure}
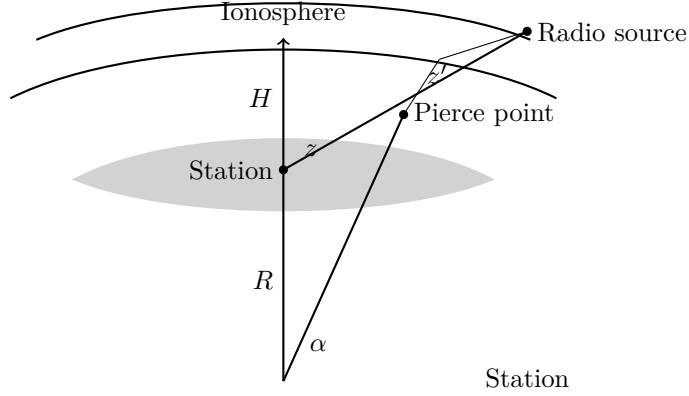

Ionospheric delay is an important and time-varying error source in VLBI observations. To correct this error, the Global Ionosphere Maps (GIMs) released by the International GNSS Service (IGS) are commonly used as a reference data source for ionospheric modelling. GIMs provide global vertical total electron content (VTEC) on a grid. In actual signal propagation, the effect of the electron content is represented by the slant total electron content (STEC) along the observation path. STEC can be converted to VTEC at the corresponding points with a projection mapping function (Eq.~\ref{eq:stec}), as illustrated in Fig.~\ref{fig:vtec-stec}.

\begin{equation}
\STEC=M(z)\,\VTEC,
\label{eq:stec}
\end{equation}
where $M(z)$ is the projection function at the ionospheric pierce point, $z$ is the receiver zenith distance, and $\VTEC$ and $\STEC$ are the ionospheric electron contents in the zenith and signal-path directions, respectively. Under the single-layer approximation,

\begin{equation}
M(z)=\left[1-\left(\frac{R\cos z}{R+H}\right)^2\right]^{-1/2},
\label{eq:mapping}
\end{equation}
where $R$ is the Earth radius and $H$ is the height of the ionospheric thin shell. The angle between the receiver and the pierce point and the zenith distance at the pierce point are denoted by $\alpha$ and $z'$, respectively, in Fig.~\ref{fig:vtec-stec}.

The ionosphere is one of the most active regions of Earth's atmosphere and consists of large numbers of free electrons and ions \citep{ref4}. Its state strongly depends on solar activity. Solar flares, coronal mass ejections, and the sunspot cycle modulate ionospheric electron density through high-energy particle flows and electromagnetic radiation. During solar-activity maxima, a single solar flare can increase the local total electron content (TEC) several-fold within tens of minutes, producing an ionospheric storm and substantially increasing the time variation of ionospheric delay. Even during quiet solar periods, the day--night contrast and latitudinal inhomogeneity of solar radiation produce regional ionospheric gradients.

Conventional VLBI systems commonly use combined S/X dual-frequency observations to remove first-order ionospheric delay. In some practical situations, however, such as observations of weak sources or observations in frequency bands with strong radio-frequency interference, only single-frequency VLBI observations are possible. Ionospheric delay in single-frequency VLBI is then generally corrected with an external ionospheric model. Existing global GNSS ionospheric models have limitations: post-processed ionospheric products are more accurate but are released with a delay and therefore cannot readily meet the needs of real-time or quasi-real-time \UT estimation; predictive models are limited by their model accuracy and cannot accurately represent real-time regional ionospheric variations and gradients, restricting their use in single-frequency VLBI \UT measurements.

To address these problems, we use GNSS data from the Jilin and Kashi VLBI stations of the National Time Service Center, Chinese Academy of Sciences, to construct a single-station regional ionospheric model. We apply the model to correct ionospheric delay in VLBI data and compare it with two external ionospheric corrections: the ionospheric predictive model from the Center for Orbit Determination in Europe (CODE) and the global post-processed ionospheric model from the Jet Propulsion Laboratory (JPL). We evaluate the VLBI delay-measurement accuracy and the final \UT solution accuracy obtained with the different correction strategies.

\section{Data Processing and Methods}
This study addresses the ionospheric-correction requirements of single-frequency VLBI \UT measurements. We use VLBI observations and co-located GNSS data from the Jilin--Kashi baseline collected from November 2024 to June 2025. A single-station regional ionospheric model is constructed with the Bernese GNSS Software and combined with two global ionospheric models. VLBI data processing, systematic-error correction, and \UT parameter estimation are then performed, and the effects of the different ionospheric correction strategies on measurement accuracy are evaluated.

\subsection{Construction of the Single-Station Regional Ionospheric Model}
Because GNSS provides all-weather operation, high precision, and flexible station deployment, global ionospheric correction models derived from GNSS inversion are widely used to remove ionospheric delay \citep{ref5}. To meet the needs of single-frequency VLBI \UT measurements on the Jilin--Kashi baseline, we use GNSS observations from receivers co-located with the VLBI stations and independently construct a single-station regional ionospheric model with the Bernese GNSS Software. This model is derived from single-station GNSS data and is specifically regionalized for VLBI measurements.

The Bernese GNSS Software was developed by the Astronomical Institute of the University of Bern and is one of the most widely used GNSS data-processing packages \citep{ref6}. It offers high processing precision, complete estimation functions, a clear modular architecture, robust automated batch processing, and high computational efficiency \citep{ref7}. It is also the main data-processing software used by CODE. Its GPSEST module supports ionospheric modelling and computation and can be directly used for estimating ionospheric parameters.

\subsubsection{Technical principle}
Based on electromagnetic propagation, we assume that all electrons along the signal path are concentrated in an ideal zero-thickness shell at an altitude of 450 km. Using single-station GNSS dual-frequency observations, we estimate STEC along the propagation path with carrier-phase-smoothed pseudoranges, convert STEC to VTEC above the station with a projection mapping function, and jointly estimate the coefficients of a spherical-harmonic model. The coefficients are finally converted into a regional ionospheric TEC grid file \citep{ref8}.

The GNSS dual-frequency code and carrier-phase observation equations are represented by

\begin{equation}
\begin{aligned}
P^{s}_{r,f}={}&\rho^{s}_{r}+c(\delta t_r-\delta t^{s})+T^{s}_{r}+I^{s}_{r,f}+d^{P}_{r,f}+d^{P,s}_{f}+\varepsilon^{P}_{r,f},\\
L^{s}_{r,f}={}&\rho^{s}_{r}+c(\delta t_r-\delta t^{s})+T^{s}_{r}-I^{s}_{r,f}+\lambda_f N^{s}_{r,f}+d^{L}_{r,f}+d^{L,s}_{f}+\varepsilon^{L}_{r,f}.
\end{aligned}
\label{eq:gnss}
\end{equation}
where $f$ is the satellite frequency; $r$ and $s$ identify the receiver and satellite, respectively; $P$ and $L$ are code and carrier-phase observations; $\rho$ is the geometric distance between the receiver and satellite; $c$ is the speed of light in vacuum; $\delta t_r$ and $\delta t^s$ are the receiver and satellite clock offsets; $T$ and $I$ are the tropospheric and ionospheric delays along the signal path; $d^P_{r,f}$ and $d^L_{r,f}$ are receiver hardware delays for code and carrier-phase observations; $d^{P,s}_{f}$ and $d^{L,s}_{f}$ are the corresponding satellite hardware delays; $\varepsilon^P$ and $\varepsilon^L$ are the code and carrier-phase observation noises; $\lambda_f$ is the wavelength; and $N$ is the integer ambiguity.

Using the GNSS dual-frequency code combinations $(P_1/P_2)$ and carrier-phase combinations $(L_1/L_2)$, frequency-independent errors are eliminated through linear combinations. Hatch filtering is then used to smooth the pseudorange with the carrier phase, improving the accuracy of STEC along the slant path. A single-layer projection mapping function converts STEC into the vertical VTEC above the station. Finally, under the thin-shell assumption, a spherical-harmonic model is used to construct the single-station regional ionospheric TEC model.

\subsubsection{Assessment of observation data}
\begin{figure}[tbp]
\centering
\includegraphics[width=\linewidth]{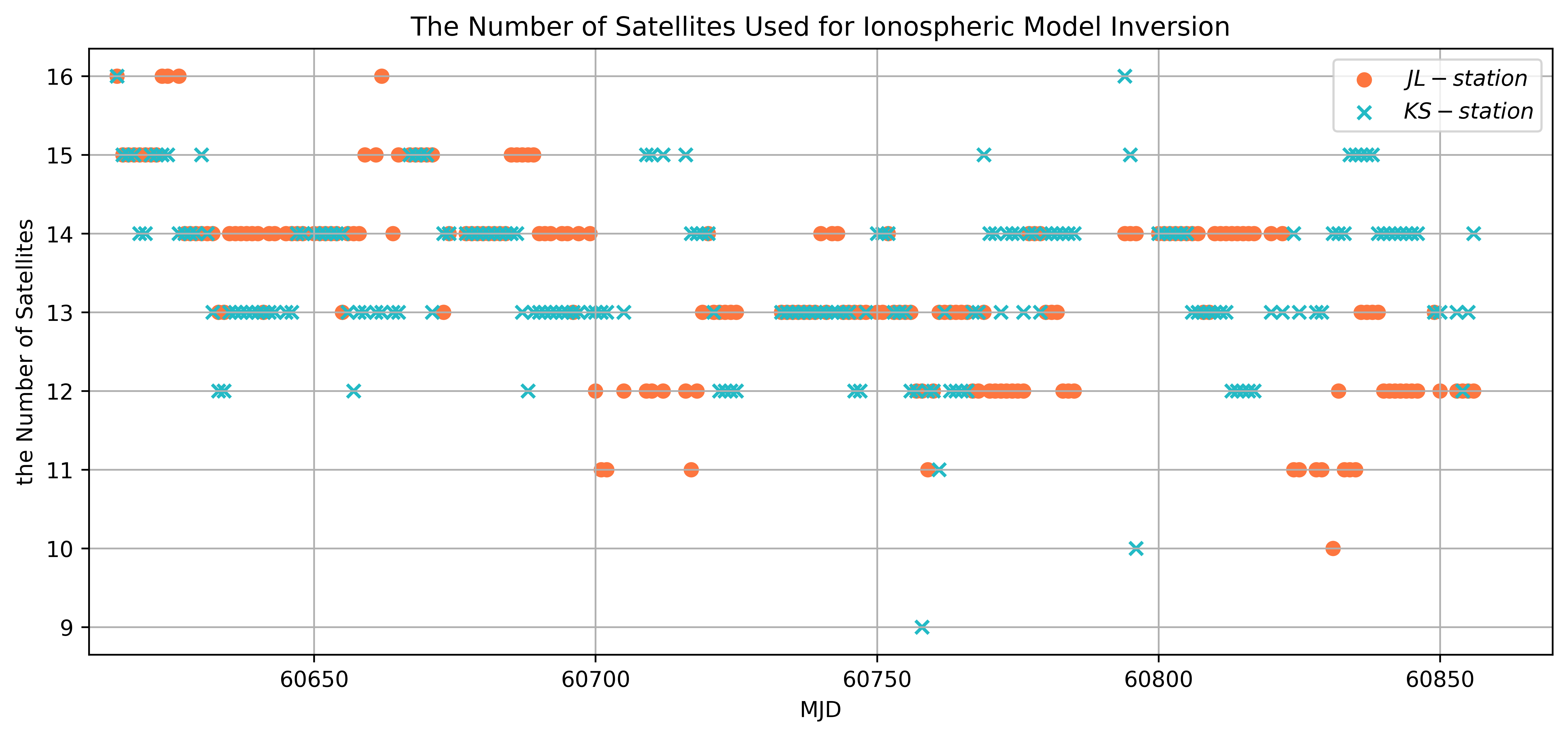}
\caption{Number of observed satellites used for the single-station regional ionospheric model inversion.}
\label{fig:sat-number}
\end{figure}

\begin{figure}[tbp]
\centering
\begin{subfigure}{0.49\linewidth}
\includegraphics[width=\linewidth]{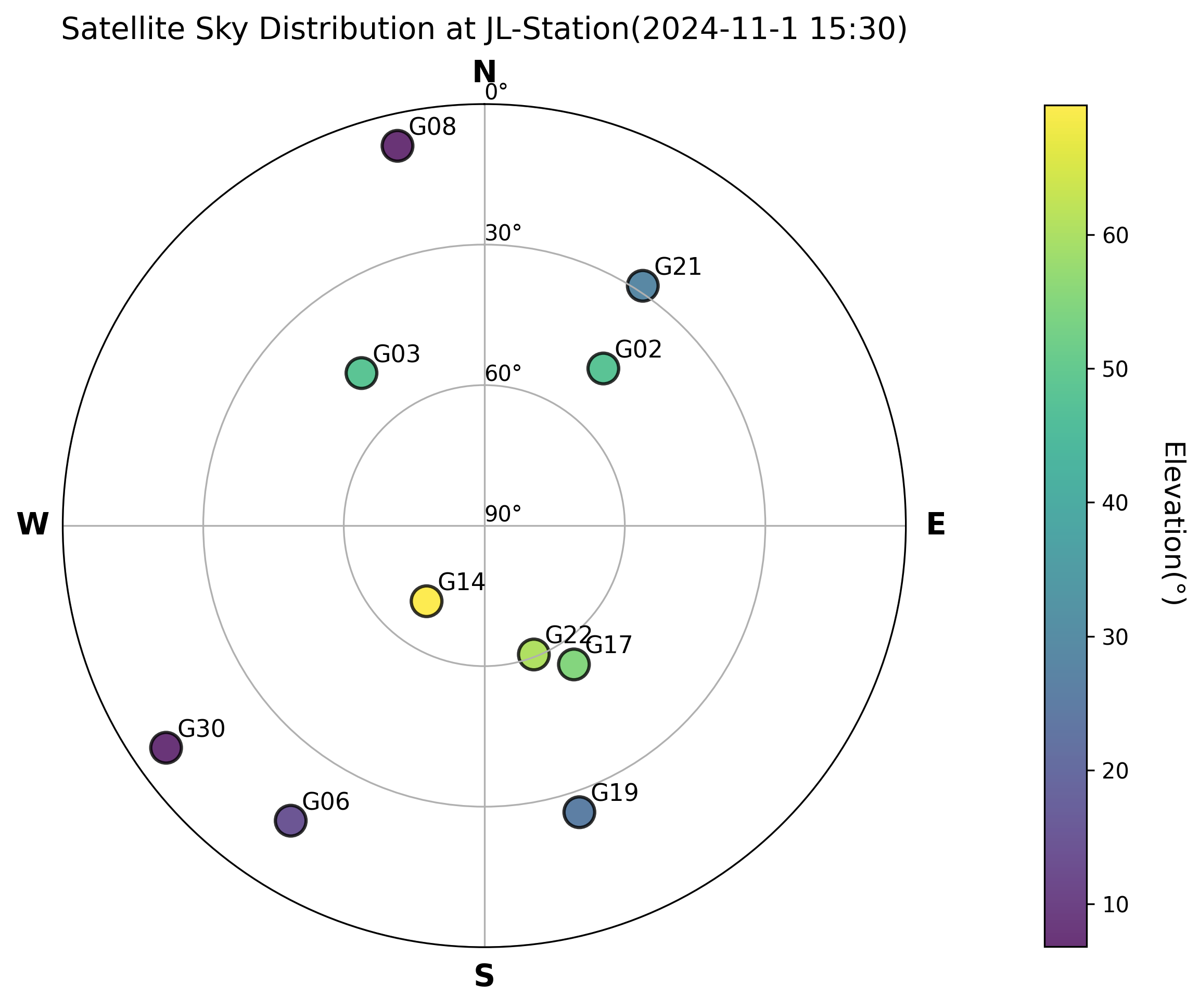}
\caption{Jilin station.}
\label{fig:sky-jl}
\end{subfigure}\hfill
\begin{subfigure}{0.49\linewidth}
\includegraphics[width=\linewidth]{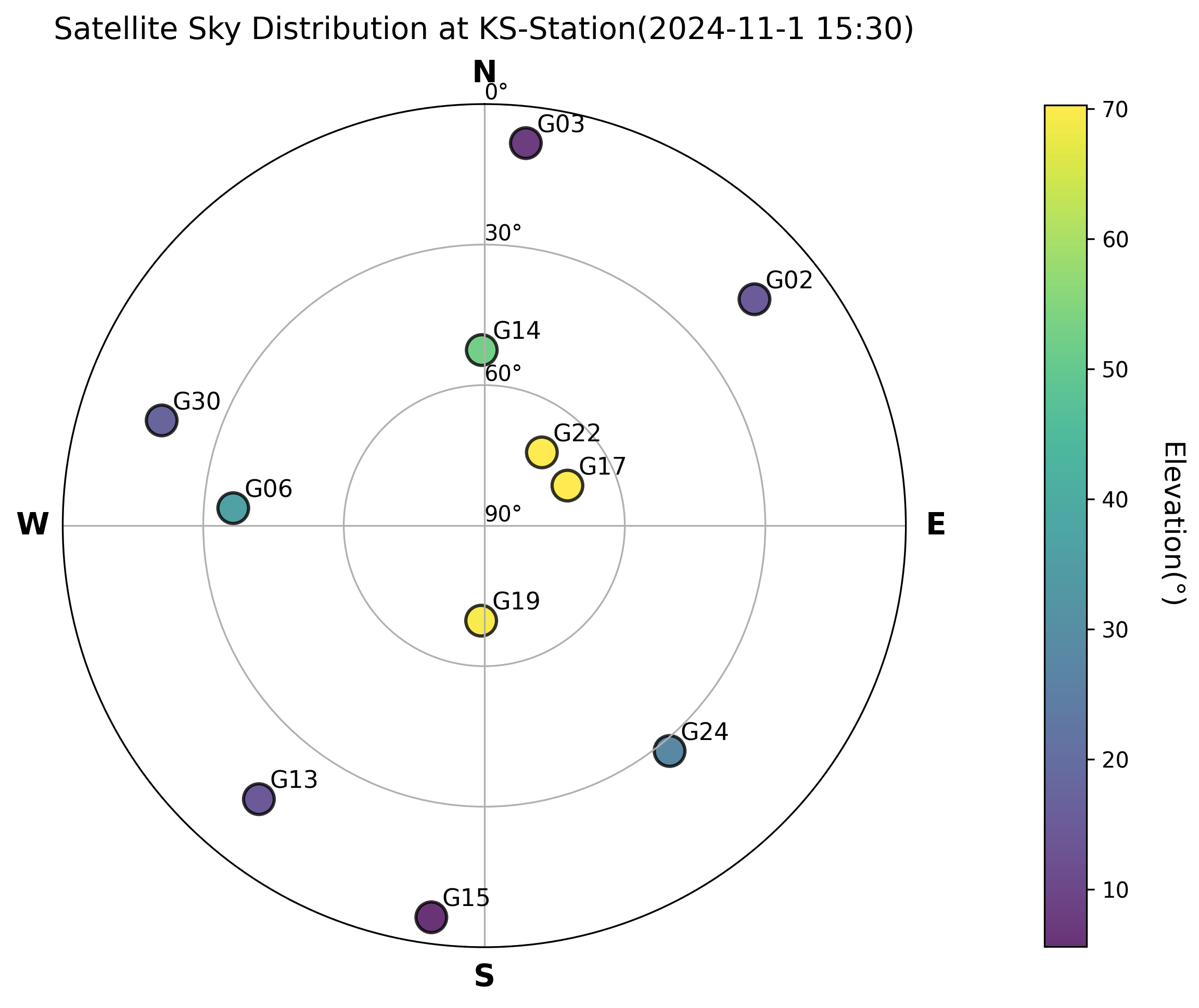}
\caption{Kashi station.}
\label{fig:sky-ks}
\end{subfigure}
\caption{Satellite sky distributions at the two stations.}
\label{fig:sky}
\end{figure}

Construction of the single-station regional ionospheric model depends on the raw GNSS observations collected by GNSS receivers co-located with the VLBI stations from November 2024 to June 2025. The data were sampled continuously at 30 s in fixed-interval mode and recorded the raw observations of satellite signals received at the stations; they are the core input for ionospheric parameter inversion. To assess the sky coverage of satellites tracked by the station GNSS receivers, we counted the GNSS satellites tracked on each day during the VLBI observing sessions and obtained the number of satellites available for the single-station regional ionospheric inversion (Fig.~\ref{fig:sat-number}). During the VLBI observing interval (15:00--17:00 UTC), the number of effectively tracked GPS satellites was 9--16, satisfying the data requirements for single-station ionospheric inversion. Further analysis of the satellite sky distribution (Figs.~\ref{fig:sky-jl} and \ref{fig:sky-ks} show the Jilin and Kashi station distributions at 15:30 UTC on 1 November 2024, respectively) indicates an approximately uniform distribution in the zenith region and provides a reliable data basis for regional ionospheric inversion.

\subsubsection{Software processing workflow}
To construct the single-station regional ionospheric model, we process the raw observations from GNSS receivers co-located with the Jilin and Kashi VLBI stations through the sequence ``data import--parameter configuration--data estimation--TEC gridding.'' The detailed workflow is as follows.

First, the software is started and the basic data are prepared. Raw GNSS observations are imported, and precise ephemerides (SP3), clock offsets (CLK), differential code biases (DCB), Earth rotation parameters (ERP), and station information (STA) files are downloaded or prepared \citep{ref9}. The observation formats are converted, and receiver-time synchronization based on code observations is performed.

During parameter configuration, the GPSEST module is set to the undifferenced mode: the difference mode is set to ZERO; the observation file containing undifferenced smoothed pseudoranges is selected; the DCB file is specified; the frequency/linear combination and the relevant strategy are selected; the BASELINE strategy is selected; and the bias parameters are set to differential code biases. In the global ionospheric-parameter panel, the parameter interval is set to 24 h, the single-layer height to 450 km, the modelling mode to STATIC, the mapping function to COSZ, and the geographical reference frame is selected; the absolute standard deviation of the coefficients is set to 10 TECU. The output panel specifies the output format and filename.

During estimation, the GPSEST module is started with the above configuration. A spherical-harmonic model and least-squares estimation are used to estimate the parameters of the single-station regional ionospheric TEC model. Multipath correction and solid-Earth-tide and ocean-tide loading corrections are also enabled to suppress non-ionospheric error sources and ensure the estimation accuracy.

After estimation, the TEC data are gridded. A grid with a spatial resolution of $2.5^{\circ}\times5^{\circ}$ is constructed to cover the signal paths between the two stations. Based on the discrete TEC parameter series, linear interpolation generates VTEC data at 2-h resolution to match the observing intervals. Kriging interpolation is used to interpolate the discrete single-station TEC data into a continuous regional-grid VTEC product. The final product is an ionospheric model in the IONosphere map EXchange (IONEX) format.

\subsection{Ionospheric Models}
To correct ionospheric delay in single-frequency VLBI observations, we use three ionospheric models; their technical parameters are listed in Table~\ref{tab:models}. CODE and JPL are both IGS analysis centres, and their ionospheric products have been validated through nearly 20 years of practice and are considered more reliable than those of other analysis centres. Their products are provided as GIMs. The global predictive and post-processed models can be downloaded from the relevant websites; predictive products can be downloaded one day in advance, whereas post-processed products lag by approximately seven days. The single-station regional model is independently estimated with the Bernese software and relies on CODE rapid satellite-orbit and clock products, so its result lags by one day.

\begin{table}[tbp]
\centering
\small
\caption{Comparison of ionospheric models. NTSC denotes the National Time Service Center.}
\label{tab:models}
\begin{tabularx}{\linewidth}{@{}lllXlX@{}}
\toprule
Product & Symbol & Source & Spatial/temporal resolution & Type & Availability \\
\midrule
Predictive ionospheric model (1 day) & Pre & CODE & $2.5^{\circ}\times5^{\circ}$, 2 h & GIM & 1 day early \\
Post-processed ionospheric model & Fin & JPL & $2.5^{\circ}\times5^{\circ}$, 2 h & GIM & Approx. 7 days late \\
Single-station regional model & Loc & NTSC & $2.5^{\circ}\times5^{\circ}$, 2 h & GIM & 1 day late \\
\bottomrule
\end{tabularx}
\end{table}

\subsection{VLBI Data Processing}
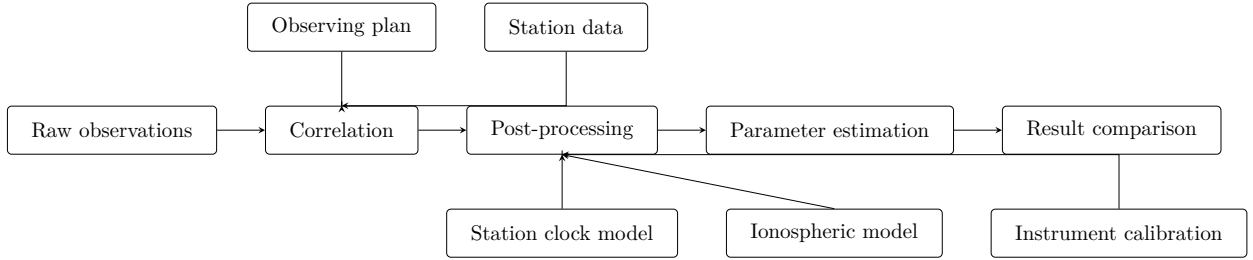
\begin{figure}[tbp]
\centering
\resizebox{\linewidth}{!}{%
\begin{tikzpicture}[
  node distance=7mm and 8mm,
  box/.style={draw, rounded corners=2pt, minimum height=8mm, inner xsep=4mm, font=\small},
  >=stealth
]
  \node[box] (raw) {Raw observations};
  \node[box, right=of raw] (cor) {Correlation};
  \node[box, right=of cor] (post) {Post-processing};
  \node[box, right=of post] (est) {Parameter estimation};
  \node[box, right=of est] (cmp) {Result comparison};
  \draw[->] (raw) -- (cor);
  \draw[->] (cor) -- (post);
  \draw[->] (post) -- (est);
  \draw[->] (est) -- (cmp);
  \node[box, above=9mm of cor] (plan) {Observing plan};
  \node[box, right=of plan] (station) {Station data};
  \draw[->] (plan.south) |- (cor.north);
  \draw[->] (station.south) |- (cor.north);
  \node[box, below=9mm of post] (clock) {Station clock model};
  \node[box, right=of clock] (iono) {Ionospheric model};
  \node[box, right=of iono] (cal) {Instrument calibration};
  \draw[->] (clock.north) |- (post.south);
  \draw[->] (iono.north) -- (post.south);
  \draw[->] (cal.north) |- (post.south);
\end{tikzpicture}
}%
\caption{VLBI data-processing and analysis workflow.}
\label{fig:vlbi-flow}
\end{figure}

We analyse single-baseline \UT measurements made with the 13-m Jilin (JL) and Kashi (KS) VLBI antennas of the National Time Service Center. The observations were conducted in the X band (8.0--8.5 GHz) from November 2024 to June 2025. Figure~\ref{fig:vlbi-flow} shows the VLBI data-processing workflow, including the input of station data and the observing schedule, raw-observation preprocessing, correlation, post-processing, parameter estimation, and result comparison. The post-processing stage introduces a station-clock model, an ionospheric model, and instrumental calibration data to correct systematic errors.

After a VLBI observing task is completed, the observation data are returned to the Xi'an data-processing centre through a network or by mailed disk for correlation and data processing. The main steps are correlation, post-processing, and parameter estimation \citep{ref10}.

\begin{enumerate}[label=(\arabic*),leftmargin=2em]
\item \textbf{Correlation.} Station coordinates in the International Terrestrial Reference Frame (ITRF), predicted initial clock offsets, and prior EOP values are introduced to calculate the theoretical-delay model. Correlation is then performed with DiFX, producing residual files containing observed delays minus theoretical delays.
\item \textbf{Post-processing.} ParselTongue and AIPS are mainly used for data inspection and editing, ionospheric, neutral-atmosphere, clock-model, and instrumental-delay corrections, fringe fitting, bandwidth synthesis, and calculation of the total baseline-delay and delay-rate series \citep{ref11}. The central task is to identify and correct systematic errors introduced by the propagation medium and instruments. A linear model is used to fit station-clock differences, and a piecewise function is used for instrumental-delay calibration. For tropospheric correction, the dry component is corrected a priori with the Saastamoinen model, whereas the wet component is estimated as a parameter. Ionospheric delay is corrected by introducing the different external models. Three rounds of fringe fitting then extract a high-precision observed-delay series from the residual data \citep{ref12}.
\item \textbf{Parameter estimation.} The theoretical and observed delay series are combined in a least-squares adjustment, and the \UT parameter is estimated with the GASV software \citep{ref13}. The result is compared with the \UT reference values released by USNO. Mean bias, RMS deviation, and mean formal error are used to evaluate measurement precision.
\end{enumerate}

Because this study compares the effects of different ionospheric models on single-frequency VLBI \UT measurements, we do not use the S/X dual-frequency method to eliminate ionospheric delay. Instead, ionospheric corrections are introduced during post-processing using the global predictive model, global post-processed model, and single-station regional model, and the results from the different strategies are compared.

\section{Results}
We compare the correction performance of the global predictive, global post-processed, and single-station regional ionospheric models at three levels: line-of-sight ionospheric delay, post-processed residual delay, and \UT parameter estimation. The single-station regional model performs best because of its local modelling advantage and substantially improves the single-frequency VLBI \UT solution.

\subsection{Comparison of Line-of-Sight Ionospheric Delays}
To compare the effects of the three ionospheric models on \UT estimation, we first compare their calculated line-of-sight ionospheric delays and then input the corrected delay data into the parameter-estimation module to analyse the final \UT results.

\begin{figure}[tbp]
\centering
\begin{subfigure}{0.49\linewidth}
\includegraphics[width=\linewidth]{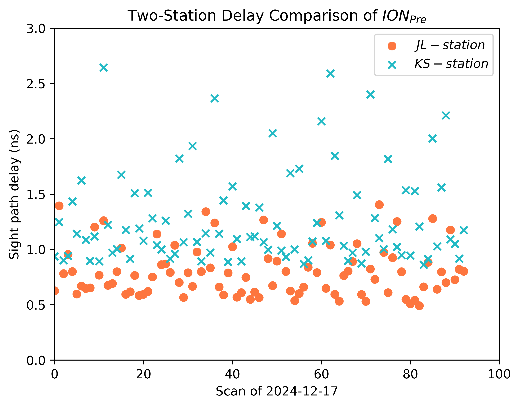}
\caption{Global predictive model.}
\end{subfigure}\hfill
\begin{subfigure}{0.49\linewidth}
\includegraphics[width=\linewidth]{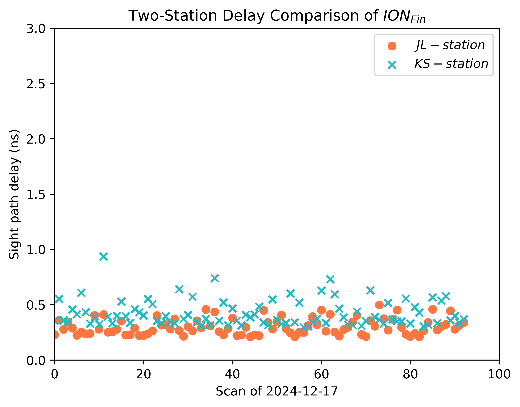}
\caption{Global post-processed model.}
\end{subfigure}\\[4pt]
\begin{subfigure}{0.49\linewidth}
\includegraphics[width=\linewidth]{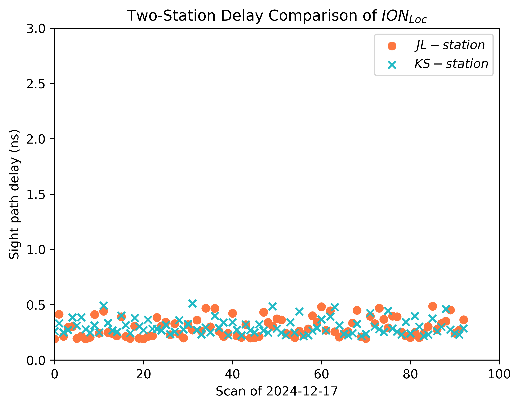}
\caption{Single-station regional model.}
\end{subfigure}
\caption{Line-of-sight ionospheric delays at the Jilin and Kashi stations calculated with different ionospheric models.}
\label{fig:los}
\end{figure}

Using the VLBI observations from the Jilin (JL) and Kashi (KS) stations between 15:00 and 17:00 UTC on 17 December 2024, we use the global predictive, global post-processed, and single-station regional models to obtain the line-of-sight ionospheric delays at the two stations by linear interpolation (Fig.~\ref{fig:los}). The horizontal axis is the VLBI SCAN sequence during the observing interval, and the vertical axis is the line-of-sight ionospheric delay in ns. Red circles represent the Jilin-station delay and blue crosses represent the Kashi-station delay.

Figure~\ref{fig:los}a shows that the delays from the global predictive model range from 0.5 to 2.5 ns, are highly dispersed, and have clearly different trends at the Jilin and Kashi stations. This reflects the limitation of the global predictive model in describing regional ionospheric spatiotemporal variations. In Figs.~\ref{fig:los}b and \ref{fig:los}c, the delays are substantially less dispersed and more concentrated, with a narrower range of 0.2--1.0 ns. This indicates that the accuracy of the single-station regional model based on GNSS data co-located with the VLBI stations is close to that of the global post-processed model.

\begin{figure}[tbp]
\centering
\includegraphics[width=0.88\linewidth]{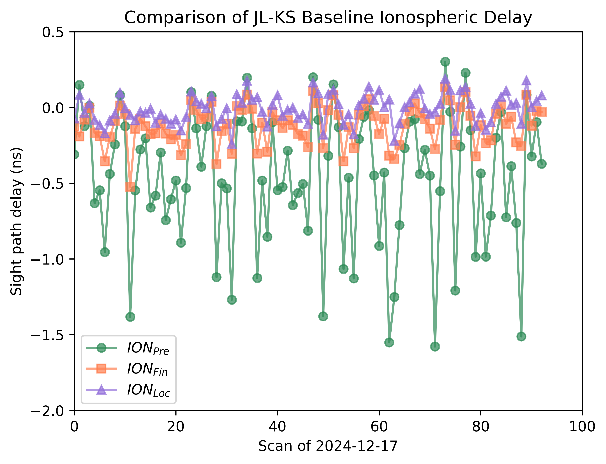}
\caption{Comparison of baseline ionospheric-delay corrections from different ionospheric models.}
\label{fig:baseline}
\end{figure}

We further interpolate the line-of-sight ionospheric delays at both stations from the same VLBI observations and calculate the baseline ionospheric correction as the Jilin-station delay minus the Kashi-station delay. The result, namely the two-station ionospheric-delay difference, is shown in Fig.~\ref{fig:baseline}. The horizontal axis is the VLBI scan sequence from 15:00 to 17:00 UTC on 17 December 2024, and the vertical axis is the Jilin--Kashi baseline ionospheric-delay correction in ns. Green circles, orange squares, and purple triangles represent the global predictive, global post-processed, and single-station regional model results, respectively.

Figure~\ref{fig:baseline} shows that the global predictive model has the largest dispersion and the largest differences from the other two models. The single-station regional model agrees closely with the global post-processed model and has a consistent variation trend. This further verifies the reliability of the single-station regional model and shows that it can replace the global post-processed model for single-frequency VLBI ionospheric correction.

\subsection{Post-Processed Residual Delays}
To verify the effectiveness of the constructed single-station regional ionospheric model for VLBI \UT measurements, we first post-process the Jilin--Kashi single-baseline observations with AIPS and then perform the comparison.

\begin{figure}[tbp]
\centering
\includegraphics[width=0.92\linewidth]{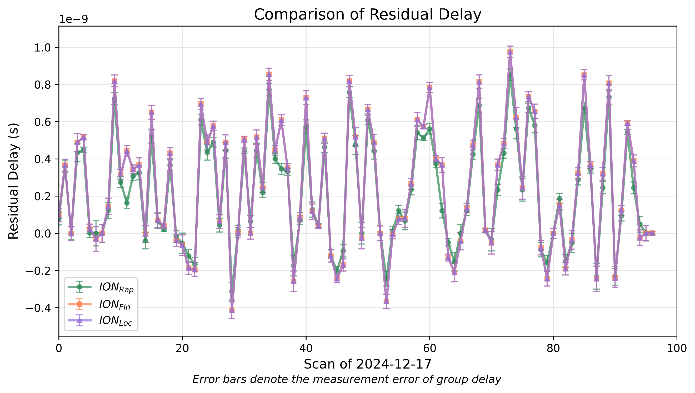}
\caption{Post-processed residual delays obtained with different ionospheric models. Error bars denote the measurement error of the group delay.}
\label{fig:residual-day}
\end{figure}

The observations from 17 December 2024 are post-processed with AIPS, including the key steps of ionospheric correction and fringe fitting. As shown in Fig.~\ref{fig:residual-day}, the residual delays after all three ionospheric corrections fall within $\pm1$ ns, directly demonstrating the effectiveness of the ionospheric-correction procedure.

The relative quality of the ionospheric models cannot be assessed directly from the range of post-processed residual delays alone, because all three corrections constrain the residual delay to the reasonable range of $\pm1$ ns. The result nevertheless verifies the reliability of the ionospheric-correction step and provides a data-quality basis for subsequent \UT parameter estimation with GASV.

\subsection{Impact on Parameter Estimation}
\begin{figure}[tbp]
\centering
\includegraphics[width=\linewidth]{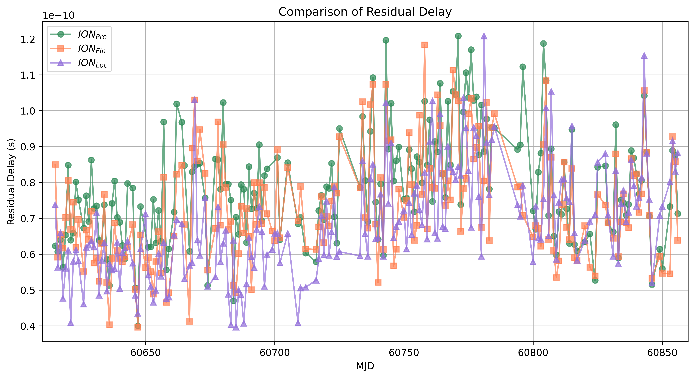}
\caption{Residual delays after parameter estimation with different ionospheric models.}
\label{fig:residual-long}
\end{figure}

During data estimation, the self-developed GASV software estimates \UT. Least-squares adjustment matches the high-precision observed-delay series from post-processing to the prior theoretical model. The magnitude of the residual delay directly reflects the agreement between the ionospheric correction model and the actual ionospheric state: a smaller residual indicates a better correction of ionospheric delay. Figure~\ref{fig:residual-long} shows the residual-delay curves obtained with different ionospheric models as a function of modified Julian date (MJD) from 1 November 2024 to 30 June 2025. The MJD range is 60615--60856. Green circles, orange squares, and purple triangles represent the results obtained with the global predictive, global post-processed, and single-station regional models, respectively.

The statistical residual-delay results allow the performance differences among the three models to be quantified. As shown in Table~\ref{tab:residual-stat}, the mean residual delays differ clearly: the mean residual from the single-station regional model is 68.6 ps, which is 4.5 ps and 10.2 ps lower than the values from the global post-processed model (73.1 ps) and global predictive model (78.8 ps), respectively. The mean residual delay of the single-station regional model is therefore lower than those of the other two models.

The dispersion of the residual delays shows the same pattern. The RMS residual of the single-station regional model is 70.2 ps, lower than 74.7 ps for the global post-processed model and 80.3 ps for the global predictive model. This indicates that the residuals from the single-station regional model are less dispersed about their mean and are more concentrated.

These results show that the single-station regional ionospheric model not only provides more accurate ionospheric-delay correction but also has better error stability, confirming its ability to resolve the fine structure of the regional ionosphere.

\begin{table}[tbp]
\centering
\caption{Statistics of residual delays after parameter estimation.}
\label{tab:residual-stat}
\begin{tabular}{lrrr}
\toprule
Ionospheric model & Mean residual (ps) & RMS residual (ps) \\
\midrule
Global predictive & 78.8 & 80.3 \\
Global post-processed & 73.1 & 74.7 \\
Single-station regional & 68.6 & 70.2 \\
\bottomrule
\end{tabular}
\end{table}

To quantify the correction performance of the different ionospheric models, we use the \UT reference values released by USNO as an independent evaluation benchmark. These values are a recognized core reference standard for EOP. Their accuracy and stability are sufficient for use as an effective benchmark and allow the results from the different ionospheric correction strategies to be compared.

\begin{figure}[tbp]
\centering
\includegraphics[width=\linewidth]{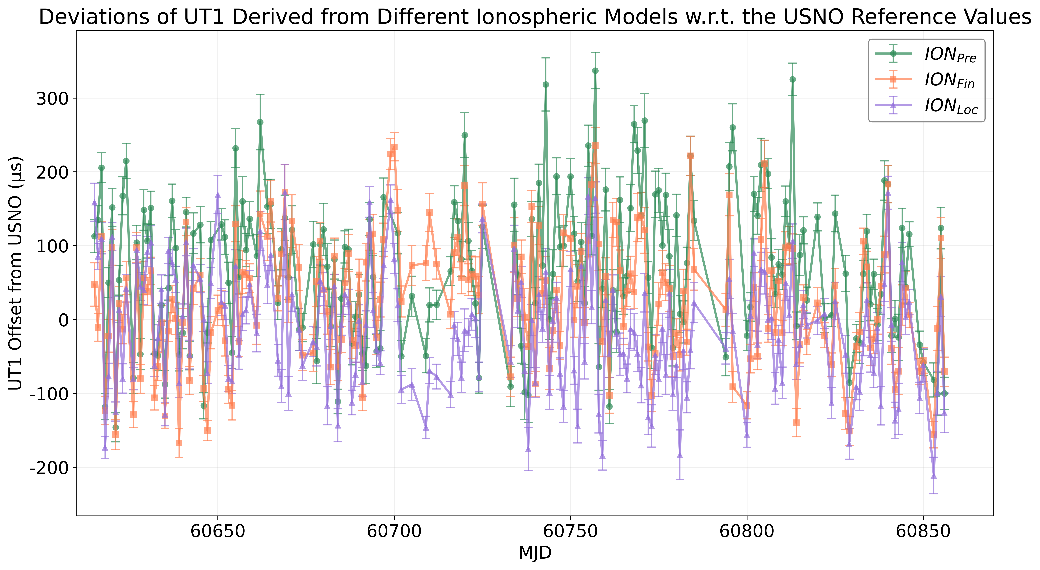}
\caption{Differences between \UT estimates obtained with different ionospheric models and the USNO reference values.}
\label{fig:ut1}
\end{figure}

Figure~\ref{fig:ut1} shows the differences between the \UT values estimated with the different ionospheric models and the USNO reference values as a function of MJD. Green circles, orange squares, and purple triangles represent the results from the global predictive, global post-processed, and single-station regional corrections, respectively. The difference is defined as the estimated \UT minus the USNO reference value, in $\mu$s. A difference closer to zero indicates a more reliable estimate.

The \UT differences obtained with the single-station regional model fluctuate only slightly around zero and are more stable than those obtained with the global predictive and global post-processed models. To quantify the comparison further, Table~\ref{tab:ut1-stat} lists the key precision indicators of the \UT solutions: mean bias, RMS deviation, and mean formal error.

\begin{table}[tbp]
\centering
\caption{Statistics of the \UT results.}
\label{tab:ut1-stat}
\begin{tabular}{llrrr}
\toprule
Ionospheric model & Source & Mean bias ($\mu$s) & RMS deviation ($\mu$s) & Mean formal error ($\mu$s) \\
\midrule
Global predictive & CODE & 70.7 & 121.4 & 23.4 \\
Global post-processed & JPL & 26.7 & 89.7 & 25.1 \\
Single-station regional & NTSC & -15.6 & 82.3 & 21.9 \\
\bottomrule
\end{tabular}
\end{table}

Table~\ref{tab:ut1-stat} shows that the \UT solution obtained with the single-station regional model has a mean bias of $-15.6\,\mu$s, an RMS deviation of $82.3\,\mu$s, and a mean formal error of $21.9\,\mu$s relative to the USNO reference values. All three indicators are better than those obtained with the global predictive model ($70.7\,\mu$s, $121.4\,\mu$s, and $23.4\,\mu$s) and the global post-processed model ($26.7\,\mu$s, $89.7\,\mu$s, and $25.1\,\mu$s).

The comparison of the different ionospheric models demonstrates that the accuracy of the \UT solution is improved by the single-station regional ionospheric model. Its central advantage is that the model is constructed from GNSS observations co-located with the Jilin and Kashi VLBI stations. It covers the complete signal-propagation path of the single Jilin--Kashi baseline and avoids the limitations of interpolation from sparsely distributed global stations. This local modelling strategy can accurately describe key regional and time-varying ionospheric features, including regional ionospheric gradients and short-term small-scale disturbances. It therefore provides single-frequency VLBI observations with ionospheric-delay corrections having a higher path match and improves the accuracy of the \UT solution.

\section{Conclusion}
This study investigates single-frequency X-band VLBI \UT measurements. Using measured VLBI and GNSS data from the Jilin and Kashi stations, we compare the performance of three ionospheric correction strategies: a global predictive model, a global post-processed model, and a single-station regional model. The main conclusions are as follows.

At the data post-processing level, the single-station regional ionospheric model performs substantially better than the global predictive model and comparably to the global post-processed model. The standard deviation of the line-of-sight ionospheric delays calculated with the single-station regional model is substantially lower than that of the global predictive model and close to that of the global post-processed model. The mean difference between the baseline ionospheric-delay correction from the single-station regional model and that from the global post-processed model is only 0.11 ns, showing that the regional model can effectively capture the fine structure of the local ionosphere.

At the parameter-estimation level, the accuracy of the ionospheric correction model directly affects the accuracy of the final \UT parameter. After correction with the independently constructed single-station regional model, the estimated \UT has the smallest systematic bias relative to the USNO reference ($-15.6\,\mu$s), the best measurement precision (RMS deviation of $82.3\,\mu$s), and good estimation stability (mean formal error of $21.9\,\mu$s). Its overall performance is better than that of the solutions obtained with the global predictive and post-processed ionospheric models.

In terms of technical and practical value, constructing a regional ionospheric model from GNSS receivers available at VLBI stations overcomes the one-to-two-week latency of global post-processed models and their inability to meet real-time needs. It also provides substantially higher accuracy than the global predictive model. The method offers an effective ionospheric-delay correction for high-precision, quasi-real-time \UT determination in single-frequency VLBI mode and is important for improving the real-time monitoring capability of EOP.

\bibliographystyle{unsrtnat}
\nocite{*}
\bibliography{references}

\end{document}